\documentclass[prb, showpacs, preprintnumbers, amsmath, amssymb, superscriptaddress, twocolumn]{revtex4}
\usepackage{graphicx, color, dcolumn, bm, natbib, amssymb, amsmath}
\usepackage[final=true]{hyperref}

\begin{document}

\title{Ferromagnetic monolayer with interfacial Dzyaloshinskii-Moriya interaction: magnon spectrum and effect of quenched disorder}

\author{Oleg I. Utesov}
\affiliation{Center for Theoretical Physics of Complex Systems, Institute for Basic Science, Daejeon 34126, Republic of Korea}
\affiliation{Petersburg Nuclear Physics Institute named by B.P.\ Konstantinov of National Research Center ``Kurchatov Institute'', Gatchina 188300, Russia}

\author{Arseny V. Syromyatnikov}
\affiliation{Petersburg Nuclear Physics Institute named by B.P.\ Konstantinov of National Research Center ``Kurchatov Institute'', Gatchina 188300, Russia}

\newcommand{\m}{\mathbf}
\newcommand{\h}{\hat}
\newcommand{\p}{\parallel}
\newcommand{\ve}{\varepsilon}
\renewcommand{\c}{\cdot}
\renewcommand{\o}{\omega}
\renewcommand{\t}{\theta}
\renewcommand{\d}{\dagger}
\newcommand{\be}{\begin{eqnarray}}
\newcommand{\ee}{\end{eqnarray}}
\newcommand{\nn}{\nonumber}

\begin{abstract}

We discuss theoretically a ferromagnetic monolayer with an interfacial Dzyaloshinskii-Moriya interaction (iDMI) and a small axial anisotropy. It is shown that the system has a long-period cycloid magnetic order slightly distorted by the anisotropy whose modulation vector $\bf k$ can have several orientations. We find that due to iDMI-induced umklapp terms in the Hamiltonian, the spectrum of long-wavelength magnons 
is essentially anisotropic: it is linear and quadratic for momenta directed along and perpendicular to $\bf k$, respectively.  Due to such a quasi-1D spectrum, the temperature correction to the mean spin value has a power-law singularity hampering the magnetic ordering at $T\ne0$. We demonstrate that the umklapps lead to a peculiar band structure of the magnon spectrum similar to electronic bands in solids.
We discuss also the effect of vacancies and defect bonds on system properties at $T=0$. Owing to the quasi-1D character of the spectrum, the distortion of the cycloid structure by a single defect bond is described by the field of 1D electric dipole. As a result, even infinitesimal concentrations of defects $c$ destroy the long-range order and establish a short-range order whose correlation length shows 
a power-law dependence on $1/c$. Our findings should also be applicable to ultra-thin films with strong enough iDMI.

\end{abstract}

\maketitle

\section{Introduction}

Noncollinear magnetic structures and corresponding materials are the subject of extensive research at the present time. A variety of spin orderings, including topologically-nontrivial ones (see, e.g., Refs.~\cite{bogdanov1989,bogdanov1994,muhlbauer2009,gobel2021beyond}), can be stabilized, in particular, by antisymmetric Dzyaloshinskii-Moriya interaction~\cite{dzyaloshinsky1958,moriya1960,dzyaloshinskii1964}. Importantly, skyrmions and other types of magnetic solitons have promising applications in memory devices~\cite{fert2013} and even in unconventional computing (see, e.g., Refs.~\cite{sai2021,raab2022brownian,lee2023perspective} and references therein).

It is well known \cite{fert1980,crepieux1998dzyaloshinsky} that due to the inversion symmetry breaking an interfacial Dzyaloshinskii-Moriya  interaction (iDMI) emerges at interfaces between magnetic and non-magnetic layers (the latter are usually made of heavy metals with large spin-orbit coupling). The iDMI can stabilize magnetic cycloid spirals. If the cycloid ordering in the system is equally possible with several modulation vectors $\bf k$, N\'eel skyrmions can emerge~\cite{fert1990magnetic,bogdanov2001chiral,bogdanov2002magnetic,bode2007chiral,ferriani2008,meckler2009}. Probably the most studied effect of the iDMI on elementary excitations is related to the asymmetry of magnon spectra in the collinear state arising in strong enough magnetic field~\cite{udvardi2009,zakeri2010,kostylev2014interface,stashkevich2015}.
This can be used for experimental quantification of the iDMI strength (see Ref.~\cite{kuepferling2023} and references therein). To the best of our knowledge, the spin dynamics of systems with iDMI in noncollinear states has not been discussed in detail. It is well known that  noncollinear spin structures possess gapless low-energy magnons (sometimes called phasons
) with a linear dispersion in phases having a single vector of magnetic ordering $\bf k$ (see, e.g., Ref.~\cite{zheludev1999magnetic}) and a quadratic spectrum in a triple-$\bf k$ skyrmion lattice~\cite{tatara2014phasons}.

In the present paper, we analytically consider a ferromagnetic monolayer with the iDMI and a small axial anisotropy. This system has a long-period cycloid magnetic structure slightly distorted by the anisotropy. This model should also be relevant to ultra-thin films. We show that the spectrum of long-wavelength elementary excitations in this system is not a simple linear function of the momentum. Due to iDMI-induced umklapp terms in the Hamiltonian, the spectrum acquires the form
\be \label{specIntro}
  \ve_\m{q} \propto \sqrt{ k^2 q^2_\p   + g^2  q^4_\perp }, \quad q \ll k\ll1.
\ee
where $\m{k}$ is the cycloid modulation vector, $g\sim1$ is a certain constant, $q_\p$ and $q_\perp$ are components of the momentum $\m{q}$ along and perpendicular to $\m{k}$, respectively. Noteworthy, a similar form of the spectrum was observed in Ref.~\cite{Maleyev2006} in cubic B20 helimagnets. One can see from Eq.~\eqref{specIntro} that the spectrum is essentially anisotropic and quasi-one-dimensional (quasi-1D) with the speed of spin waves being zero in the direction perpendicular to $\m{k}$. 

We obtain that the umklapp terms in the Hamiltonian lead to strong hybridization of magnons with momenta $q_\| = \pm k/2, \pm k, \pm 3k/2,\dots$ and, as a consequence, to the emergence of gaps in the spectrum inside the Brillouin zone. Then, we observe a peculiar magnon band structure similar to electronic bands in solids. We show that magnons from different bands produce anomalies in dynamical spin susceptibilities at a given momentum. Thus, our prediction of the magnon band structure can be verified experimentally.

We show that, despite the quasi-1D form of the spectrum, the correction to the mean spin value diverges only at finite temperatures. However, this divergence is power-law in contrast to the logarithmic one in isotropic 2D systems. We demonstrate that the correlation volume of the short-range order at finite $T$ should be essentially elongated along $\m{k}$.

As interfaces cannot be ideal in practice, we discuss the effect of quenched disorder and consider in detail two particular examples of defects. The first one is related to disorder in neighboring nonmagnetic layers which should lead to variations of parameters of the spin Hamiltonian (the so-called bond disorder). The second type is represented by vacancies in the magnetic material. 
We obtain that the cycloid distortion caused by a single impurity is described by an electrostatic field produced by a complex of charges (a dipole or several dipoles) placed at the defect location. At a finite concentration $c$ of impurities, the long-range order is suppressed and a short-range order emerges due to the long-range influence of defects on the ground state. A similar effect was recently obtained in 2D frustrated helimagnets~\cite{santanu2020}, where the correlation length of the short-range order depends exponentially on $1/c$. However, the influence of defects on the cycloid order in the present model is stronger at large distances due to the quasi-1D long-wavelength magnon spectrum. As a result, we obtain a power-law dependence on $1/c$ of the correlation length of the impurity-induced short-range order.

The rest of the present paper is organized as follows. In Sec.~\ref{SecModel}, we introduce the model under consideration and the technique which eventually leads us to the bosonic Hamiltonian suitable for the derivation of the classical magnon spectrum. The latter is discussed in Sec.~\ref{SecSpec}, where we also consider the effect of finite temperature on the magnetic order. Quenched disorder produced by vacancies and/or defect bonds is discussed in Sec.~\ref{SecDis}. We present our conclusions in Sec.~\ref{SecConc}.

\section{Model and technique}
\label{SecModel}

\begin{figure}
  \centering
  \includegraphics[width=5cm]{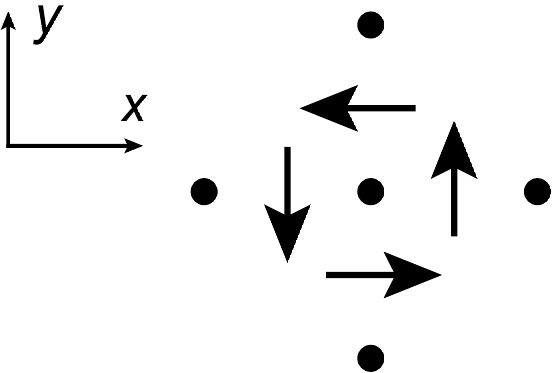}\\
  \caption{Square-lattice magnetic monolayer which is assumed to be sandwiched between different non-magnetic materials. Due to the broken inversion symmetry, the interfacial Dzyaloshinskii-Moriya interaction (iDMI) arises. Directions of iDMI vectors among the central ion and its nearest neighbors are shown by arrows.
 }\label{figDMI}
\end{figure}

\subsection{Spin interactions and Hamiltonian}

We start with the spin Hamiltonian on the square lattice and set the lattice parameter equal to unity. The model to be considered includes the dominant ferromagnetic exchange coupling $\mathcal{H}_{\textrm{EX}}$, interfacial Dzyaloshinskii-Moriya interaction (iDMI) $\mathcal{H}_{\textrm{DM}}$, and single-ion anisotropy $\mathcal{H}_{\textrm{AN}}$  which is assumed to be much smaller than the iDMI:
\be \label{ham1}
  \mathcal{H}_0 &=& \mathcal{H}_{\textrm{EX}} + \mathcal{H}_{\textrm{DM}} + \mathcal{H}_{\textrm{AN}}, \\
  \mathcal{H}_{\textrm{EX}} &=& -\frac{1}{2} \sum_{ij} J_{ij} (\m{S}_i \cdot \m{S}_j), \nn \\
  \mathcal{H}_{\textrm{DM}} &=& \frac12 \sum_{ij} \m{D}_{ij} \cdot \left[ \m{S}_i \times \m{S}_j \right], \nn \\
  \mathcal{H}_{\textrm{AN}} &=& - B \sum_i \left(S^z_i\right)^2, \nn
\ee
where $\m{D}_{ij}=-\m{D}_{ji}$. We consider both cases of the easy-axis anisotropy $B>0$ and the easy-plane one $B<0$. Notice that in the long-wavelength limit, there is an effective contribution of the easy-plane type to $\mathcal{H}_{\textrm{AN}}$ from the magnetodipolar interaction~\cite{maleev1976}. We neglect other symmetry-allowed anisotropic interactions (e.g., the anisotropic exchange and the square anisotropy) which are normally much smaller in real systems than those presented above. However, these interactions can lead to important subtle effects in spiral magnets, e.g., preferable directions for the modulation vector (see, e.g.,  Ref.~\cite{bak1980} and below).

After the Fourier transform
\be \label{fourier1}
  \m{S}_j = \frac{1}{\sqrt{N}} \sum_j \m{S}_\m{q} e^{i \m{q} \cdot \m{R}_j},
\ee
where $N$ is the number of the lattice sites and $\m{q}$ is the two-dimensional wave vector, we have from Eqs.~\eqref{ham1} 
\be \label{ham2}
  \mathcal{H}_{\textrm{EX}} &=& - \frac12 \sum_{\m{q}} J_{\m{q}} \m{S}_\m{q} \cdot \m{S}_{-\m{q}}, \nn \\
  \mathcal{H}_{\textrm{DM}} &=& \frac12 \sum_{\m{q}} \m{D}_{\m{q}} 
	\cdot \left[\m{S}_\m{q} \times \m{S}_{-\m{q}} \right], \\
  \mathcal{H}_{\textrm{AN}} &=& - B \sum_\m{q} S^z_\m{q} S^z_{-\m{q}}. \nn
\ee
For the Fourier transform of the exchange interaction, we use below its long-wavelength expansion
\be \label{exch}
  J_\m{q} \cong J_\m{0} - \frac{A q^2}{S}.
\ee

The general expression for the iDMI Fourier transform reads as
\be \label{DMI1}
  \m{D}_\m{q} = - i \sum_\m{b} D_\m{b} \sin{(\m{q} \cdot \m{b})} [\hat{z} \times \m{b}],
\ee
where the summation includes all the lattice bonds $\m{b}$ of a certain lattice site. In the case of high-symmetry lattices (e.g., square or triangular ones), Eq.~\eqref{DMI1} can be universally rewritten in the long-wavelength limit as
$
  \m{D}_\m{q} \cong - i D_0 [\hat{z} \times \m{q}].
$
When iDMI is considered between the nearest-neighbor spins only, $D_0 = 2 D$ for the square-lattice monolayer (see Fig.~\ref{figDMI}) and $D_0 = 3 D$ for the triangular-lattice one. For definiteness, below we consider the square-lattice monolayer and use the following explicit form of iDMI in particular calculations:
\be \label{DMI2}
  \m{D}_\m{q} \cong - 2 i D [\hat{z} \times \m{q}].
\ee
There are also higher order in $\m{q}$ corrections to $\m{D}_\m{q}$ in Eq.~\eqref{DMI1}, which in the nearest-neighbor model on the square lattice have the form
\be\label{deltad}
  \delta \m{D}_\m{q} = - \frac{i D}{3} \left( \hat{x} q^3_y - \hat{y} q^3_x \right).
\ee
We show below that despite their smallness they can play an important role in both static and dynamical properties of the model.

\subsection{Ground-state energy and magnetic ordering}
\label{secor}

\subsubsection{$B=0$}

Let us neglect the anisotropy for a start. We assume a planar ground-state spin ordering, adopt Kaplan's description of helical structures~\cite{Kaplan1961, Maleyev2006}, and introduce the following local Cartesian basis at each lattice site $\bf R$:
\be \label{kaplan}
  \hat{\zeta}_\m{R} &=& \hat{u} \cos(\m{k} \m{R}+\phi) + \hat{b}  \sin(\m{k} \m{R}+\phi), \nn \\
  \hat{\eta}_\m{R} &=& - \hat{u} \sin(\m{k} \m{R}+\phi) + \hat{b}  \cos(\m{k} \m{R}+\phi), \\
  \hat{\xi}_\m{R} &=& \hat{c}, \nn
\ee
where $\phi$ is an arbitrary phase, $\m{k}$ is the momentum of the magnetic ordering, $\hat{u}$, $ \hat{b}$, and $ \hat{c}$ are unit vectors, and ${[\hat{u} \times \hat{b}] = \hat{c}}$. We obtain below a stable spectrum of excitations that justifies our assumption about the coplanar spiral ordering. It is convenient to use auxiliary vectors
\be\label{AA}
  \m{C} = \frac{\hat{u} - i \hat{b}}{2}, \quad \m{C}^* = \frac{\hat{u} + i \hat{b}}{2}
\ee
which have the following properties:
\be
  \m{C} \cdot \m{C} =0, \quad \m{C} \cdot \m{C}^* =\frac12, \quad \left[ \m{C} \times \m{C}^* \right] = \frac{i \hat{c}}{2}, \\
  \left[ \m{C} \times \hat{c} \right] = - i \m{C}, \quad \left[ \m{C}^* \times \hat{c} \right] =  i \m{C}^*.
\ee
Then, one obtains from Eqs.~\eqref{fourier1}, \eqref{kaplan}, and \eqref{AA} 
\be\label{ssq}
  \m{S}_\m{q} &=& \m{C} e^{i\phi} (S^\zeta_{\m{q} -\m{k}} + i S^\eta_{\m{q} -\m{k}} ) +  \m{C}^* e^{-i\phi} (S^\zeta_{\m{q} + \m{k}} - i S^\eta_{\m{q} +\m{k}} ) \nn\\
  &&{}+ \hat{c} S^\xi_\m{q}.
\ee

Substituting Eq.~\eqref{ssq} to Eqs.~\eqref{ham2} and putting $S^\zeta_\m{q} = S \sqrt{N} \delta_{\m{q},\m{0}}$, $S^\eta_\m{q}=S^\xi_\m{q}=0$,
one obtains the classical energy of the system
\be \label{en1}
  \frac{E}{S^2 N} = - \frac12 J_\m{k} - D \m{k} \cdot [\hat{z} \times \hat{c}].
\ee
Eq.~\eqref{en1} should be minimized with respect to $\bf k$ and orientations of $\hat{u}$, $\hat{b}$, and $\hat{c}$ with respect to the global coordinate frame $xyz$ shown in Fig.~\ref{figDMI}. Since $J_\m{k}$ depends on $|\m{k}|$ only [see Eq.~\eqref{exch}], the iDMI energy in Eq.~\eqref{en1} is minimized when
\be\label{kzc}
  \m{k} \uparrow \uparrow D [\hat{z} \times \hat{c}] 
\ee
and $|[\hat{z} \times \hat{c}]|=1$. Thus, $\hat{c}$ should lie in the $xy$ plane and $\hat{b}$ can be directed along $z$ axis so that
\be\label{bzu}
  \hat{b} = \frac{D}{|D|}\hat{z}, \quad \hat{u} = \frac{\m{k}}{k}.
\ee
As $\bf k$ lies in the plane of spin rotations ($ub$), the obtained spin ordering is a cycloid. Different signs of $D$ correspond to the right- and to the left-handed cycloids. Also using Eq.~\eqref{exch}, one obtains as a result from Eq.~\eqref{en1} 
$$
   \frac{E}{S^2 N} = - \frac12 J_\m{0} + \frac{A}{2S} k^2 - |D| k.
$$
Minimization of this expression gives
\be \label{mod}
  k = S \frac{|D|}{A}
\ee
and
\be
   \frac{E}{S^2N} = - \frac12 J_\m{0} - \frac{A}{2S} k^2.
\ee

Notice that the $\m{k}$ direction remains arbitrary in the $xy$ plane. However, tiny corrections \eqref{deltad} from iDMI produce the additional term in Eq.~\eqref{en1}
\be\label{de}
\frac{\delta E}{S^2N} = \frac{|D|}{6k}\left(k_x^4+k_y^4\right)
\ee
which set the $\bf k$ direction to be parallel to square diagonals. The correction~\eqref{de} is of the fourth order in spin-orbit coupling [see Eq.~\eqref{mod}]. However, there are other small spin interactions in this order in real systems that we do not consider here and which can compete with correction~\eqref{de} and fix other $\bf k$ directions (e.g., square edges).

\subsubsection{$B\ne0$}

The small single-ion anisotropy gives the correction $-B ( \hat u_z^2 + \hat b_z^2 )/2$ to the right-hand side of Eq.~\eqref{en1}. It is easy to realize that taking into account this term does not change Eqs.~\eqref{bzu} and \eqref{mod} if $|B|\ll |D|k$ and the ground-state energy acquires the form
\be \label{en2}
   \frac{E}{S^2N} = - \frac12 J_\m{0} - \frac{A}{2S} k^2 - \frac12 B.
\ee
However, we show below that the axial anisotropy distorts the cycloid and gives rise to an additional correction to Eq.~\eqref{en2} proportional to $B^2 / A k^2$. In our calculations, we concentrate on the regime of a perturbative treatment of the anisotropy which is possible when $S |B| \ll Ak^2$.

The quantity $A k^2$ is the characteristic energy scale of the cycloid structure. For example, the field of transition to the fully polarized phase reads approximately as $ A k^2/g \mu_B$ (see, e.g., Ref.~\cite{maleyev2019}). 
Upon $|B|$ increase, the cycloid becomes further distorted according to the chiral soliton lattice scenario (see, e.g., Refs.~\cite{dzyal1964III, Izyumov1984, kishine2015}) and the first-order transition to the collinear spin structure takes place at $B=B_c$, where
\be\label{bc}
  |B_c| = \frac{\pi^2}{8} \frac {A k^2}{S}.
\ee

\subsection{Transformation of spin Hamiltonian}

Substituting Eq.~\eqref{ssq} to Eqs.~\eqref{ham2}, we obtain for the exchange interaction
\be \label{hex}
  \mathcal{H}_\textrm{EX} &=& - \frac12 \sum_\m{q} \Biggl[ J_\m{q} S^\xi_\m{q} S^\xi_{-\m{q}} + J_{\m{q},\m{k}} \left( S^\zeta_\m{q} S^\zeta_{-\m{q}} + S^\eta_\m{q} S^\eta_{-\m{q}} \right) \nn \\ &&{}+ i N_{\m{q},\m{k}} \left( S^\eta_\m{q} S^\zeta_{-\m{q}}  - S^\zeta_\m{q} S^\eta_{-\m{q}} \right) \Biggr],
\ee
where
\be \label{Jqk}
  J_{\m{q},\m{k}} &=& \frac{J_{\m{q} + \m{k}} + J_{\m{q} - \m{k}} }{2} \cong J_\m{0} - \frac{A (q^2 + k^2)}{S} , \\ \label{Nqk}
  N_{\m{q},\m{k}} &=& \frac{J_{\m{q} + \m{k}} - J_{\m{q} - \m{k}} }{2} \cong - \frac{2 A (\m{q} \cdot \m{k})}{S}.
\ee

Contributions to the Hamiltonian from iDMI can be divided into two classes: direct and umklapp terms. The former provide terms in the bilinear part of the bosonic Hamiltonian in which the total momentum of created and annihilated magnons is conserved, whereas the total momentum changes by $\pm \m{k}$ in umklapp terms. After some calculations, we obtain for the direct terms
\be \label{hdmid}
  \mathcal{H}^d_\textrm{DM} &=& - D k \sum_\m{q} \left( S^\zeta_\m{q} S^\zeta_{-\m{q}} + S^\eta_\m{q} S^\eta_{-\m{q}} \right) \nn \\
  &&{}- \frac{i D}{k} \sum_\m{q} (\m{q} \cdot \m{k}) \left( S^\eta_\m{q} S^\zeta_{-\m{q}}  - S^\zeta_\m{q} S^\eta_{-\m{q}} \right).
\ee
The umklapps can be written as
\be \label{hdmiu}
  \mathcal{H}^u_\textrm{DM} &=&   - i D \sum_\m{q} (\m{q} \cdot \hat{c}) \, \left(S^\eta_{\m{q}+\m{k}}+S^\eta_{\m{q}-\m{k}}\right) S^\xi_{-\m{q}}    \\
  && + D \sum_\m{q} (\m{q} \cdot \hat{c}) \left( S^\zeta_{\m{q}+\m{k}} S^\xi_{-\m{q}}  + S^\xi_\m{q} S^\zeta_{-\m{q}-\m{k}} \right). \nn
\ee

The anisotropy also contains direct and umklapp terms. However, the latter violate the momentum conservation law by $\pm 2 \m{k}$. Their explicit forms are the following:
\be
\label{hdan}
  \mathcal{H}^d_\textrm{AN} &=&  - \frac{B}{2} \sum_\m{q} 
  \left( S^\zeta_\m{q} S^\zeta_{-\m{q}} + S^\eta_\m{q} S^\eta_{-\m{q}}  \right),\\
\label{huan}
  \mathcal{H}^u_\textrm{AN} &=&  
    \frac{B}{4} \sum_\m{q} \Bigl( S^\zeta_{\m{q}-\m{k}} S^\zeta_{-\m{q}-\m{k}} - S^\eta_{\m{q}-\m{k}} S^\eta_{-\m{q}-\m{k}}  \\ 
  && {} + S^\zeta_{\m{q}+\m{k}} S^\zeta_{-\m{q}+\m{k}} - S^\eta_{\m{q}+\m{k}} S^\eta_{-\m{q}+\m{k}} 
  \Bigr) \nn \\ 
  && {} + 
  \frac{i B}{4} \sum_\m{q}  \Bigl( S^\zeta_{\m{q}-\m{k}} S^\eta_{-\m{q}-\m{k}} + S^\eta_{\m{q}-\m{k}} S^\zeta_{-\m{q}-\m{k}} \nn \\&& 
  {}- S^\zeta_{\m{q}+\m{k}} S^\eta_{-\m{q}+\m{k}} - S^\eta_{\m{q}+\m{k}} S^\zeta_{-\m{q}+\m{k}} 
  \Bigr). \nn
\ee

\subsection{Bosonic Hamiltonian at $B=0$}

Using the Holstein-Primakoff representation,~\cite{Holstein1940} we write the spin components in the local basis~\eqref{kaplan} as follows:
\be \label{SpinRep}
  &&S^\zeta_\m{q} = S \sqrt{N} \delta_{\m{q},\m{0}} - \frac{1}{\sqrt{N}}\sum_{\m{q}_1} a^\d_{\m{q}_1} a_{\m{q}_1 + \m{q}}, \nn \\
  &&S^\eta_\m{q} \cong \sqrt{\frac{S}{2}} \Biggl[a^\d_{-\m{q}} + a_\m{q}  \\ 
  &&{} - \frac{1}{4SN} \sum_{\m{q}_1,\m{q}_2} \left( a^\d_{\m{q}_1} a_{\m{q}_2}  a_{\m{q} + \m{q}_1 - \m{q}_2} + a^\d_{\m{q}_1} a^\d_{\m{q}_2}  a_{\m{q} + \m{q}_1 + \m{q}_2} \right) \Biggr], \nn \\
  &&S^\xi_\m{q} \cong i \sqrt{\frac{S}{2}} \Biggl[ a^\d_{-\m{q}} - a_\m{q}  \nn \\
   &&{}+ \frac{1}{4SN} \sum_{\m{q}_1,\m{q}_2} \left( a^\d_{\m{q}_1} a_{\m{q}_2}  a_{\m{q} + \m{q}_1 - \m{q}_2} - a^\d_{\m{q}_1} a^\d_{\m{q}_2}  a_{\m{q} + \m{q}_1 + \m{q}_2} \right) \Biggr]. \nn
\ee
We need magnon interaction terms in the Hamiltonian for an accurate calculation of the magnon spectrum.

Substituting Eqs.~\eqref{SpinRep} to Eqs.~\eqref{hex}, \eqref{hdmid}, \eqref{hdmiu}, \eqref{hdan}, and \eqref{huan} one obtains the bosonic analog of the spin Hamiltonian~\eqref{ham1}. In this Hamiltonian, terms without Bose operators give the system classical energy which has been discussed in Sec.~\ref{secor}. As we treat the anisotropy as a perturbation, let us discuss first the bosonic Hamiltonian at $B=0$. It can be checked straightforwardly that terms in the Hamiltonian linear in Bose operators vanish or cancel each other at $k$ given by Eq.~\eqref{mod}. For the bilinear part of the Hamiltonian, we have
\be \label{bilinear}
  \mathcal{H}^{(2)} = \sum_{\m{q}} \Biggl[ {\cal E}_\m{q} a^\d_{\m{q}} a_{ \m{q}}  + {\cal B}_\m{q} \frac{a^\d_{\m{q}} a^\d_{ -\m{q}} + a_{\m{q}} a_{- \m{q}}}{2} \\
  {}+ {\cal C}_\m{q} \left( a^\d_{\m{q}} a_{ \m{q}-\m{k}} + a^\d_{\m{q}} a_{ \m{q}+\m{k}}\right) \Biggr], \nn
\ee
where
\be \label{EBC}
  {\cal E}_\m{q} &=& A q^2 + \frac{A k^2}{2}, \nn \\
  {\cal B}_\m{q} &=& - \frac{A k^2}{2}, \\
  {\cal C}_\m{q} &=& A k (\m{q} \cdot \hat{c}). \nn
\ee
For reasons discussed below, the most important terms describing magnon interaction are triple terms originating from the iDMI umklapps~\eqref{hdmiu}
\be \label{hdmiu2}
  \mathcal{H}^{(3)} &=& i D \sqrt{\frac{S}{2N}} \sum_{\m{q},\m{q_1}} (\m{q} \cdot \hat{c}) \\ 
  && {}\times a^\d_{\m{q}_1} \left( a^\d_{\m{q}+\m{k}} - a_{-\m{q}-\m{k}} - a^\d_{\m{q}-\m{k}} + a_{\m{k}-\m{q}} \right) a_{\m{q}_1+\m{q}} .\nn
\ee

\subsection{Bosonic Hamiltonian at $B\ne0$}

Both direct \eqref{hdan} and umklapp \eqref{huan} terms of the anisotropy provide the following contributions to the bilinear part of the Hamiltonian:
\be
\label{hand}
 \mathcal{H}^{d(2)}_{\textrm{AN}} &=& \frac{S B}{2} \sum_{\m{q}} \left[ a^+_\m{q} a_\m{q} - \frac{a^+_{\m{q}} a^+_{ -\m{q}} + a_{\m{q}} a_{- \m{q}}}{2}\right], \\
 \label{hanu}
  \mathcal{H}^{u(2)}_{\textrm{AN}} &=&  -\frac{3 S B}{4} \sum_\m{q} \left( a^+_{\m{q}} a_{ \m{q}- 2 \m{k}} + a^+_{\m{q}} a_{ \m{q}+2 \m{k}} \right) \nn \\
  && - \frac{S B}{4} \sum_{\m{q}} \frac{ a_{\m{q}} a_{2 \m{k} - \m{q}} + a_{\m{q}} a_{- 2 \m{k} -  \m{q}}}{2}  \\
  &&-  \frac{S B}{4} \sum_{\m{q}} \frac{a^+_{\m{q}} a^+_{ 2 \m{k} - \m{q}} + a^+_{\m{q}} a^+_{ -2 \m{k} - \m{q}}}{2}. \nn
\ee

Importantly, terms linear in Bose operators do not cancel each other in $\mathcal{H}^u_\textrm{AN}$. We have for them from Eq.~\eqref{huan}
\be \label{hanlin}
   \mathcal{H}^{(1)}_{\textrm{AN}} &=&  i \left( \frac{S}{2} \right)^{3/2} B \sqrt{N} \left( a_{-2\m{k}} - a_{2\m{k}} + a^+_{2\m{k}}  - a^+_{-2\m{k}} \right). \nn \\
\ee
These terms should be eliminated from the Hamiltonian by the following shift in operators:
\be \label{shift1}
  a_{2\m{k}} &\Rightarrow& a_{2\m{k}} + \rho_+ e^{i \varphi_+}, \quad a_{-2\m{k}} \Rightarrow a_{-2\m{k}} + \rho_- e^{i \varphi_-}, \\
  a^+_{2\m{k}} &\Rightarrow& a^+_{2\m{k}} + \rho_+ e^{-i \varphi_+}, \quad a^+_{-2\m{k}} \Rightarrow a^+_{-2\m{k}} + \rho_- e^{- i \varphi_-}, \nn
\ee
where $\rho_\pm$ and $\varphi_\pm$ are real numbers. Terms in the Hamiltonian linear in Bose operators should cancel each other after this shift. Thus, the following set of equations should hold:
\be \label{set1}
  - i \left( \frac{S}{2} \right)^{3/2} B \sqrt{N} + {\cal E}_{2 \m{k}} \rho_+ e^{-i \varphi_+} + {\cal B}_{2\m{k}} \rho_- e^{i \varphi_-} &=& 0,  \nn \\
  \\
  i \left( \frac{S}{2} \right)^{3/2} B \sqrt{N}  + {\cal E}_{2 \m{k}} \rho_- e^{-i \varphi_-} + {\cal B}_{2\m{k}} \rho_+ e^{i \varphi_+} &=& 0, \nn 
\ee
where ${\cal E}_{2 \m{k}}$ and ${\cal B}_{2\m{k}}$ are given by Eqs.~\eqref{EBC}. The solution of Eqs.~\eqref{set1} reads as
\be \label{shift2}
  \rho_\pm = \left( \frac{S}{2} \right)^{3/2} 
  \frac{B}{4 A k^2} \sqrt{N}, 
  \quad 
  e^{-i \varphi_+} = e^{i \varphi_-} = i.
\ee
Then, one has from Eq.~\eqref{shift1}
\be \label{shift3}
  \langle a_{2\m{k}} \rangle =  \langle a^+_{-2\m{k}} \rangle = - i \rho_+, \quad  \langle a_{-2\m{k}} \rangle =  \langle a^+_{2\m{k}} \rangle = i \rho_+.
\ee
It is pertinent to discuss three main consequences of ``condensation''~\eqref{shift3} of the bare bosons.

First, shift \eqref{shift1} leads to the correction to the classical energy~\eqref{en2} having the form
\be
  \frac{\delta E}{S^2N} = - \frac{S B^2}{16 A k^2} + {\cal O}(B^3)
\ee
which results in the following renormalization of the modulation vector in the leading order in $B$ (cf. Ref.~\cite{izyumov1983neutron}):
\be
  k^\prime = k - \frac{S^2 B^2}{8 A^2 k^3},
\ee
where $k$ is given by Eq.~\eqref{mod}. This correction is ${\cal O}(B^2)$ and it will be omitted below. Equality $k'=0$ provides a correct estimation of the critical value $|B_c|\sim Ak^2/S$ given by Eq.~\eqref{bc}.

Second, Eqs.~\eqref{shift1} lead to the elliptical distortion and to the third harmonic in the cycloid structure (see, e.g., Ref.~\cite{utesov2018}) which can be obtained from Eqs.~\eqref{ssq} and \eqref{SpinRep} with the result
\be\label{sr}
\mathbf{S}_\m{R} &=& S\left[ \hat{u} \left(1-\frac{SB}{8Ak^2}\right) \cos (\mathbf{k} \mathbf{R} + \phi) \right. \\
&& {}+ \hat{b} \left(1+\frac{SB}{8Ak^2}\right) \sin (\mathbf{k} \mathbf{R} + \phi)  \nn\\
&& \left.{}+\frac{SB}{8Ak^2}\left(\hat{u} \cos (3 \mathbf{k} \mathbf{R} + \phi) + \hat{b} \sin (3 \mathbf{k} \mathbf{R} + \phi)\right) \right] .\nn
\ee

Third, $\pm \m{k}$ and $\pm 3 \m{k}$ umklapps linear in $B$ arise in the bilinear part of the Hamiltonian from Eq.~\eqref{hdmiu2} after shift~\eqref{shift1}. Umklapps $\pm \m{k}$ read as
\be \label{hucorr}
  \delta \mathcal{H}^u = - A k \frac{S B}{8 A k^2} \sum_\m{q} (\m{q} \cdot \hat{c}) \left( a^+_{\m{q}} a_{ \m{q}-\m{k}} + a^+_{\m{q}} a_{ \m{q}+\m{k}}\right),
\ee
whereas umklapps $\pm 3\m{k}$ are negligible within the accuracy of our further calculations in the leading order in $B$.

\section{Magnon spectrum}
\label{SecSpec}

\subsection{General discussion}


As it is seen from Eqs.~\eqref{bilinear}, \eqref{EBC}, \eqref{hand}, \eqref{hanu}, and \eqref{hucorr}, umklapp terms and anomalous terms ($a^\d_{\m{q}} a^\d_{ -\m{q}}$ and $a_{\m{q}} a_{- \m{q}}$) in the Hamiltonian contain small parameters: either $k$ or $B$. Then, it is evident from Eqs.~\eqref{bilinear} and \eqref{EBC} that the spectrum of short-wavelength excitations with $q \gg k$ almost coincides with the magnon dispersion of an isotropic ferromagnet:
\be
  \ve_\m{q} \cong A q^2, \quad q \gg k,
\ee
where $A$ plays the role of the spin wave stiffness.

The opposite limiting case of $q \ll k$ is much more subtle because umklapps mix all the magnons with momenta $\m{q} + n \m{k}$, where $n$ is an integer. However, it is clear from the above consideration that the classical spectrum should be gapless because the phase $\phi$ introduced in Eq.~\eqref{kaplan} remains arbitrary in our discussion of the ground state [a simultaneous rotation of all spins in the $ub$ plane by changing $\phi$ in Eq.~\eqref{sr} costs no energy]. 

To find the spectrum at finite $q \ll k$, we introduce normal and anomalous Green's functions
\be\label{gf}
G_n(q) &=& \langle a_{\m{q}+n\m{k}}, a^\d_\m{q} \rangle_\omega,\nn\\
\overline{G}_n(q) &=& \langle a^\d_{\m{q}+n\m{k}}, a_\m{q} \rangle_\omega,\nn\\
F_n^\d(q) &=& \langle a^\d_{-\m{q}-n\m{k}}, a^\d_\m{q} \rangle_\omega,\\
F_n(q) &=& \langle a_{-\m{q}-n\m{k}}, a_\m{q} \rangle_\omega,\nn
\ee
where $n$ is integer and $q=(\omega, {\bf q})$. There is an infinite set of Dyson equations for them even in the harmonic (classical) approximation in which one has to take into account only the bilinear part of the Hamiltonian given by the sum of Eqs.~\eqref{bilinear}, \eqref{hand}, \eqref{hanu}, and \eqref{hucorr}. To find Green's functions and the magnon spectrum, one has to truncate this set of equations at some $n_{tr}$. We find below that $n_{tr}=2$ is sufficient to obtain results in the leading orders in small parameters $Aq^2$ and $B$. It happens due to the smallness of the umklapp terms in the Hamiltonian at $q \ll k$. 

The set of the Dyson equations for $n_{tr}=2$ has the form
\be
\hat{M}\left(
\begin{array}{c}
G_{-2}(q)\\
F_{-2}(q)\\
G_{-1}(q)\\
F_{-1}(q)\\
G_0(q)\\
F_0(q)\\
G_1(q)\\
F_1(q)\\
G_2(q)\\
F_2(q)
\end{array}
\right)
=
\left(
\begin{array}{c}
0\\
0\\
0\\
0\\
1\\
0\\
0\\
0\\
0\\
0
\end{array}
\right),
\ee
where the matrix $\hat{M}$ reads as
\begin{widetext}
\be\label{m}
\hat{M} = \left(
    \begin{array}{cccccccccc}
      \omega-\tilde{{\cal E}}_{\m{q} - 2 \m{k}}  &  -\tilde{{\cal B}}_{\m{q} - 2 \m{k}} & -\tilde{{\cal C}}_\m{q} & 0 & \frac{3SB}{4} & \frac{SB}{4} & 0 & 0 & 0 & 0 \\
      \tilde{{\cal B}}_{\m{q} - 2 \m{k}} & \omega+\tilde{{\cal E}}_{\m{q} - 2 \m{k}}  & 0 & -\tilde{{\cal C}}_\m{q} & -\frac{SB}{4} & -\frac{3SB}{4} & 0 & 0 & 0 & 0 \\
      -\tilde{{\cal C}}_\m{q} & 0 & \omega-\tilde{{\cal E}}_{\m{q} - \m{k}}  & -\tilde{{\cal B}}_{\m{q} - \m{k}} & -\tilde{{\cal C}}_\m{q} & 0 & \frac{3SB}{4} & \frac{SB}{4} & 0 & 0 \\
      0 & -\tilde{{\cal C}}_\m{q} & \tilde{{\cal B}}_{\m{q} - \m{k}} & \omega+\tilde{{\cal E}}_{\m{q} - \m{k}}  & 0 & -\tilde{{\cal C}}_\m{q} & -\frac{SB}{4} & -\frac{3SB}{4} & 0 & 0 \\
      \frac{3SB}{4} & \frac{SB}{4} & -\tilde{{\cal C}}_\m{q} & 0 & \omega-\tilde{{\cal E}}_{\m{q}}  & -\tilde{{\cal B}}_{\m{q}} & -\tilde{{\cal C}}_\m{q} & 0 & \frac{3SB}{4} & \frac{SB}{4} \\
      -\frac{SB}{4} & -\frac{3SB}{4} & 0 & -\tilde{{\cal C}}_\m{q} & \tilde{{\cal B}}_{\m{q}}  &  \omega+\tilde{{\cal E}}_{\m{q}}  & 0 & -\tilde{{\cal C}}_\m{q} & -\frac{SB}{4} & -\frac{3SB}{4} \\
      0 & 0 & \frac{3SB}{4} & \frac{SB}{4} & -\tilde{{\cal C}}_\m{q} & 0 & \omega-\tilde{{\cal E}}_{\m{q} + \m{k}} & -\tilde{{\cal B}}_{\m{q} + \m{k}} & -\tilde{{\cal C}}_\m{q} & 0 \\
      0 & 0 & -\frac{SB}{4} & -\frac{3SB}{4} & 0 & -\tilde{{\cal C}}_\m{q} &  \tilde{{\cal B}}_{\m{q} + \m{k}} & \omega+\tilde{{\cal E}}_{\m{q} + \m{k}}  & 0 & -\tilde{{\cal C}}_\m{q} \\
      0 & 0 & 0 & 0 & \frac{3SB}{4} & \frac{SB}{4} & -\tilde{{\cal C}}_\m{q} & 0 & \omega-\tilde{{\cal E}}_{\m{q} + 2 \m{k}}  &  -\tilde{{\cal B}}_{\m{q} + 2\m{k}} \\
      0 & 0 & 0 & 0 & -\frac{SB}{4} & -\frac{3SB}{4} & 0 & -\tilde{{\cal C}}_\m{q} & \tilde{{\cal B}}_{\m{q} + 2\m{k}} & \omega+\tilde{{\cal E}}_{\m{q} - 2 \m{k}}  \\
    \end{array}
  \right). \nn \\
\ee
\end{widetext}
In Eq.~\eqref{m} we omit self-energy parts (quantum and thermal corrections), parameters have the form
\be
  \tilde{{\cal E}}_\m{q} &=& A q^2 + \frac{A k^2}{2} + \frac{S B}{2}, \\
  \label{b}
  \tilde{{\cal B}}_\m{q} &=& - \frac{A k^2}{2} - \frac{S B}{2}, \\
  \tilde{{\cal C}}_\m{q} &=& A k (\m{q} \cdot \hat{c}) \left(1 - \frac{SB}{8Ak^2}\right),
\ee
and we use the equality $\tilde{{\cal C}}_\m{q} = \tilde{{\cal C}}_{\m{q} + n \m{k}}$ which follows from Eq.~\eqref{kzc}. 

The spectrum is given by the roots of the equation
\be
\det \hat{M}=0,
\ee
where $\hat{M}$ is given by Eq.~\eqref{m}. One obtains in the leading order in $Aq^2$ and $B$
\be \label{spec3}
  \ve_\m{q} &=& \sqrt{(Ak^2 + SB) A q^2_\p + A^2 f(\m{q})},\\
  q_\perp &=& (\m{q} \cdot \hat{c}),\\
  q_\| &=& (\m{q} \cdot \m{k})/k.
\ee
Here we take into account Eq.~\eqref{bzu} and introduce
\be
  f(\m{q}) = q^4_\p - 4q^2_\p q^2_\perp + \frac38q^4_\perp,
\ee
which is of the fourth order in $q$. Evidently, the term with $f(\m{q})$ in Eq.~\eqref{spec3} becomes important at small $q_\p$ only so that we can rewrite Eq.~\eqref{spec3} in a more compact form
\be \label{spec4}
  \ve_\m{q} = \sqrt{(Ak^2 + SB) A q^2_\p + \frac{3}{8} A^2 q^4_\perp}.
\ee
Almost the same result was obtained in Ref.~\cite{Maleyev2006} for cubic B20 helimagnets (without the uniaxial anisotropy). Straightforward calculations with $n_{tr}=3$ do not change Eq.~\eqref{spec4} in the leading order in $Aq^2$ and $B$.


The main feature of the obtained spectrum \eqref{spec4} is that it is essentially anisotropic. The speed of spin waves is proportional to $|\cos{\varphi}|$, where $\varphi$ is the angle between momentum $\m{q}$ and the cycloid vector $\m{k}$. In the limit of $\varphi \to \pi/2$, the spectrum becomes quadratic, $\ve_\m{q} \propto A q^2_\perp$. This happens due to the cancellation of terms proportional to $(Ak^2+SB)q^2_\perp$ in Eq.~\eqref{spec4} so that higher order in $q_\perp$ terms come into play. That is why one has to reconsider the above derivation based on Eqs.~\eqref{exch} and \eqref{DMI2} taking into account further $O(q^4)$ terms in $J_\m{q}$ and $O(q^3)$ terms \eqref{deltad} in iDMI. The result for the square-lattice monolayer has the form
\be \label{spec5}
  \ve_\m{q} = \sqrt{(Ak^2 + SB) A q^2_\p + g^2 A^2 q^4_\perp},
\ee
which differs from Eq.~\eqref{spec4} by $g^2=13/24$. Besides, small terms of the order of $A^2 k^4 q^2_\perp$ also emerge in ${\ve}^2_\m{q}$ which are of the further order in the spin-orbit coupling and which can come into play at $q_\perp\ll k^2$ only. However, careful accounting for such terms requires consideration of other anisotropic interactions in further orders in the spin-orbit coupling which exist in real systems and which we omit in the present study. Thus, we can safely use Eq.~\eqref{spec5} below, bearing in mind that it is valid at $k \gg q \gtrsim k^2$. 

\subsection{Magnon band structure}

An interesting feature of the considered system is the appearance of the Bragg (resonant) scattering due to the umklapps which leads to the emergence of magnon bands similar to electronic bands in solids (see, e.g., Ref.~\cite{Kittel}). For example, magnons with $\m{q} = \pm \m{k}$ are no longer propagating particles due to the anisotropy-induced double umklapps. This effect can be revealed using a two-state effective Hamiltonian which can be represented by a $4 \times 4$ matrix. Corresponding eigenstates are symmetric and antisymmetric combinations of ``bare'' magnons and their energies are
\be\label{band1}
  \ve_\pm = \sqrt{2} A k^2 \left[ 1 + \frac{2 S B }{8 A k^2} \pm \frac{5 S |B| }{8 A k^2} \right].
\ee
Then, the gap proportional to $|B|$ opens at $\bf q = \pm k$. Its linearity in the anisotropy constant is a consequence of the resonant mixing of degenerate magnon states. Noteworthy, this splitting can be probed using the spin resonance technique because $\bf S_0$ contains $S^\eta_{\pm\bf k}$ [see Eq.~\eqref{ssq}]. As a result, the dynamical two-spin correlator built on $\bf S_0$ contains $G_0(\omega, {\pm\bf k})$ and $F_0(\omega, {\pm\bf k})$.

\begin{figure}
  \centering
  \includegraphics[width=8cm]{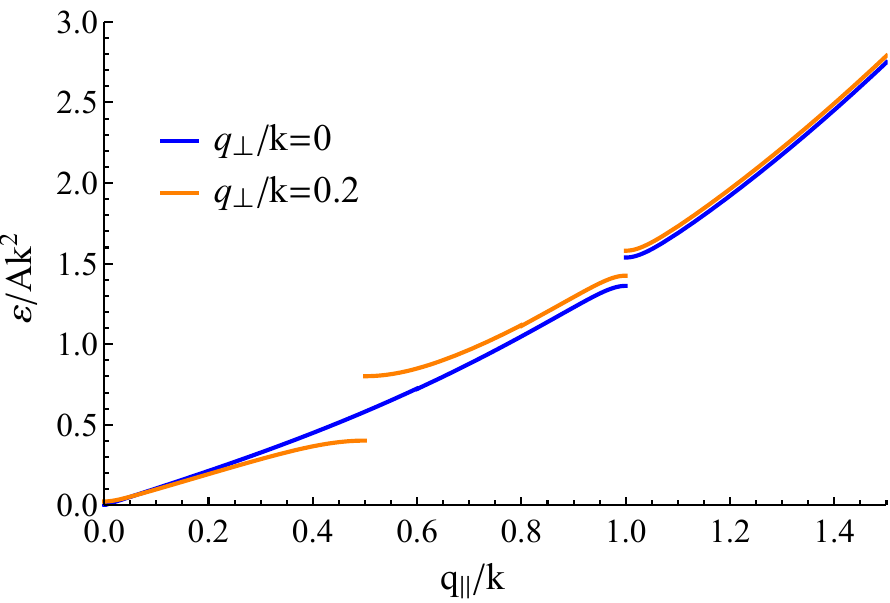}\\
  \caption{Magnon energy as a function of the momentum component $q_\|$ parallel to the cycloid modulation vector $\bf k$ for two values of the transverse component of the momentum $q_\perp \ll k$. In both cases, the gap at $q_\p = k$ is due to the single-ion anisotropy, whereas at $q_\perp \neq 0$ iDMI induces the gap at $q_\p = k/2$ [see Eqs.~\eqref{band1} and \eqref{band2}]. Here, the anisotropy parameter $S B=0.1 Ak^2$. 
  \label{FigSpec}}
\end{figure}

A similar effect arises for $q_\p = \pm k/2$ at $q_\perp \ll k$. In this case, the gap linear in $q_\perp$ opens up due to single umklapps induced by the iDMI so that at $q_\p = \pm k/2$ one has
\be\label{band2}
  \ve_\pm = \frac{\sqrt{5}}{4} A k^2 + \frac{2 \sqrt{5}}{5} S B \pm A k |q_\perp| \left(1 - \frac{SB}{8 Ak^2} \right).
\ee
Fig.~\ref{FigSpec} illustrates our findings \eqref{band1} and \eqref{band2}. Noteworthy, the first and the second band has a single crossing point at $q_\p = k/2, \, q_\perp=0$.

The case of $q_\perp \sim k$ is essentially different. Indeed, at $q_\p \lesssim k$ the single umklapps are of the same order of magnitude with other coefficients in the Hamiltonian [see also Eq.~\eqref{m}]. It does not allow to truncate the system of the Dyson equations at some $n_{tr}$ in an asymptotically correct fashion. However, we solve the problem numerically and observe that the lowest magnon band can be described within $n_{tr} = 2 $ approximation with high accuracy. The reason is that the coefficients $\tilde{\cal E}_{\m{q} + n \m{k}}$ grow with $n$ approximately as $n^2$. Thus, the problem of the magnon spectrum calculation here resembles the problem of a particle hopping in the external parabolic potential (note that we discuss the ``hopping'' in the reciprocal space). The lowest energy state is localized near the potential minimum, so large ``distances'' are not important. In our case, large $n$ are not important for the lowest magnon band description. For example, if we set $n_{tr} = 3$, additional ``sites'' will be available for hopping. However, their energy is $\approx 10 A k^2$, which is much larger than the relevant energy domain for the first magnon band.

\begin{figure}
  \centering
  \includegraphics[width=8cm]{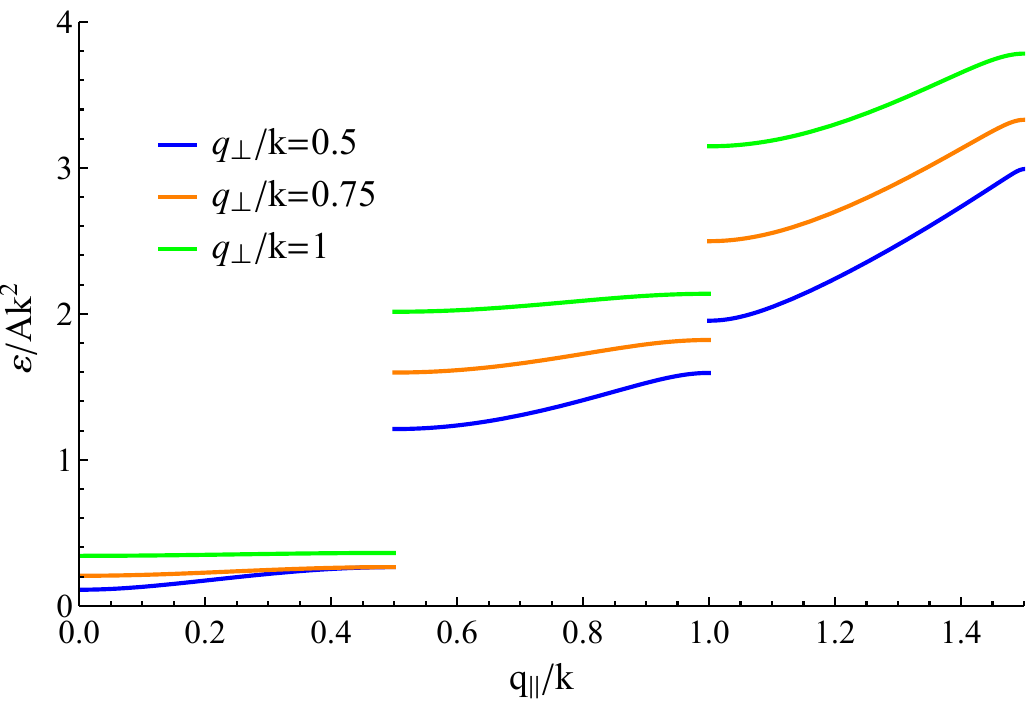}\\
  \caption{Same as Fig.~\ref{FigSpec} but for three values of $q_\perp\sim k$ and $B=0$.
  \label{FigSpec2}}
\end{figure}

Our numerical results for various $q_\perp \sim k$ values are summarized in Fig.~\ref{FigSpec2} for $B=0$ (small finite $B$ does not change the spectra qualitatively in this case). One can see that the magnon bands are almost flat at $q_\p \lesssim k$ due to large gaps caused by strong umklapps. We use $n_{tr} = 4$ for the careful description of energies in the second and third bands. It is pertinent to note that we use small-$q$ expansions of various quantities in our calculations. Thus, there is a natural ``cut-off'' for the number of bands that can be in principle discussed. Evidently, the condition  $n_{tr} k \lesssim 1$ should be satisfied.

\begin{figure}
  \centering
  \includegraphics[width=8cm]{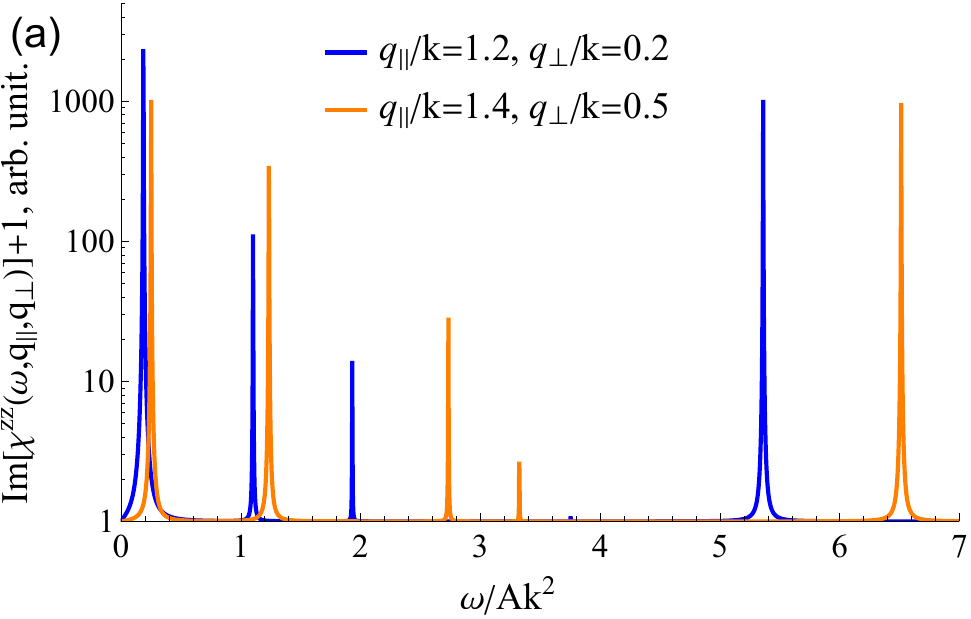}
  \hspace{1cm} \includegraphics[width=8cm]{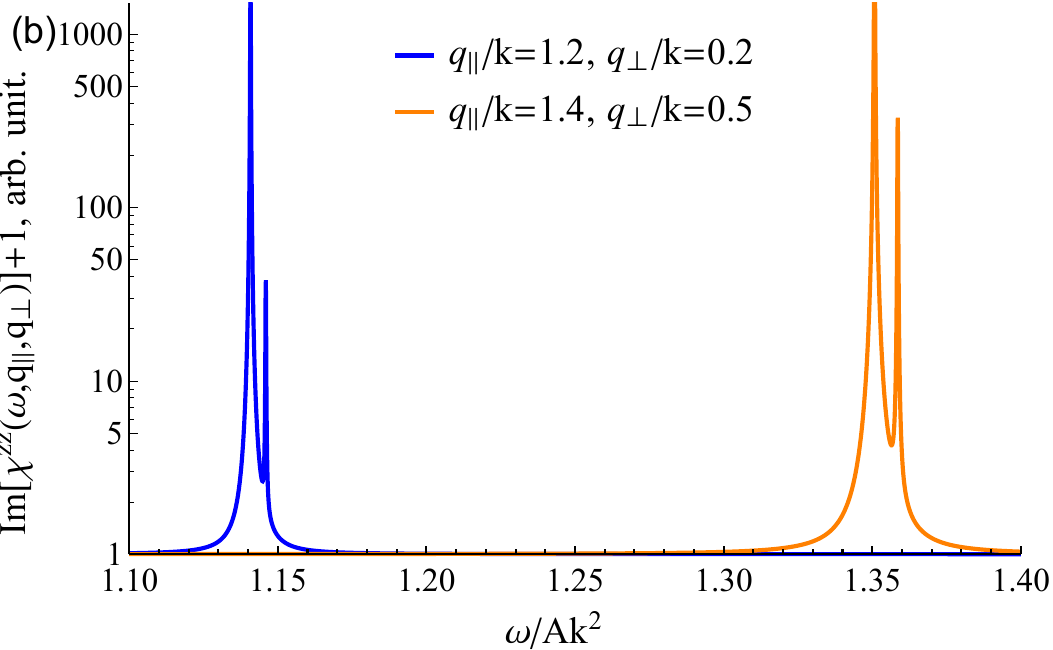}\\
  \caption{Imaginary part of the $zz$-component of the dynamical spin susceptibility calculated using Eq.~\eqref{chizz} for two momenta $\bf q$. For illustration purposes, we use the logarithmic scale, add unity to the susceptibility, and manually broaden peaks by adding $i0.001Ak^2$ to $\omega$. (a) Anisotropy value $B=0$. (b) $S B = 0.2 Ak^2$. Each peak seen in panel (a) acquires a satellite.
  \label{FigStruct2}}
\end{figure}

The peculiar magnon band structure discussed above can be obtained experimentally. The imaginary part of the dynamical spin susceptibility tensor $\chi^{\alpha\beta}(\omega,\m{q})$ is connected with the dynamical structure factor which is probed in, e.g., neutron scattering experiments. In particular, using Eqs.~\eqref{AA}, \eqref{ssq}, \eqref{bzu}, \eqref{SpinRep}, and \eqref{gf}, for $\chi^{zz}(\omega,\m{q})=-\langle S^z_{\m{q}}, S^z_{-\bf q} \rangle_\omega$  we obtain
\be\label{chizz}
  &&\chi^{zz}(\omega,\m{q}) = -\frac S8 \Bigl(
  e^{2i\phi}\Bigl[F_2(-q-k) + F_{-2}^\d(q+k) \nn\\
&&+ G_{-2}(q+k) + \overline{G}_2(-q-k)\Bigr]\\
&&+e^{-2i\phi}\Bigl[F_{-2}(-q+k) + F_{2}^\d(q-k) 
\nn\\
&&+ G_{2}(q-k) + \overline{G}_{-2}(-q+k)\Bigr]\nn\\
  &&+F_0(-q+k) + F_0^\d(q-k) + G_0(q-k) + \overline{G}_0(-q+k)\nn\\
  &&+F_0(-q-k) + F_0^\d(q+k) + G_0(q+k) + \overline{G}_0(-q-k)\nn
  \Bigr).
\ee
 In Fig.~\ref{FigStruct2} we present ${\rm Im}\chi^{zz}(\omega,\m{q})$ calculated using Eq.~\eqref{chizz} at $\phi=0$ for two momenta belonging to different regimes of relatively weak and strong umklapps influence (cf.\ Figs.~\ref{FigSpec} and~\ref{FigSpec2}). One can see that for both momenta the dynamical structure factor has numerous peaks produced by magnons from different bands. 
Noteworthy, finite $B$ leads to a double structure of each peak which is illustrated in Fig.~\ref{FigStruct2}(b).

\subsection{N\'eel temperature and correlation lengths of short-range order}

The magnon spectrum obtained above can be used to discuss the destruction of the long-range order at finite temperature $T$. The standard estimation of the N\'eel temperature comes from the equality $\delta S=\langle a^\d_i a_i\rangle=S$ which follows from the Holstein-Primakoff representation~\eqref{SpinRep} of the longitudinal spin component. One has \cite{agd} 
\be\label{ds}
  \delta S = \lim_{\tau \rightarrow -0} -T \sum^\infty_{n=-\infty} \int \frac{d^2 q}{(2 \pi)^2} G_0(i\omega_n , \m{q}) e^{-i \omega_n \tau},
\ee
where the summation over Matsubara frequencies is performed, and $G_0(q)$ is defined in Eq.~\eqref{gf}. In the leading order in small parameters, $G_0(q)$ reads as
\be\label{Gnormal}
  G_0(q) &\approx& \frac{\omega + \tilde{{\cal E}}_\m{q} }{\omega^2 - \ve^2_\m{q}},
\ee
where $\ve_\m{q}$ is given by Eq.~\eqref{spec5} and $\tilde{{\cal E}}_\m{q}\approx Ak^2/2$ at $q\ll k$.
After some calculations from Eq.~\eqref{ds} we have 
\be
  \delta S = \delta S_0 + \delta S_T,
\ee
where
\be \label{corrT0}
  \delta S_0 = \int \frac{d^2 q}{(2 \pi)^2} \frac{Ak^2 - 2 \ve_\m{q}}{4 \ve_\m{q}}
\ee
is the contribution due to ``zero-point oscillations'' and
\be \label{corrT}
  \delta S_T = \int \frac{d^2 q}{(2 \pi)^2} \frac{A k^2}{2 \ve_\m{q}} \frac{1}{e^{\ve_\m{q}/T}-1}
\ee
is the temperature correction.

In both Eq.~\eqref{corrT0} and Eq.~\eqref{corrT} infrared behavior of spectrum \eqref{spec5} is of prime importance, in which the last term plays a role only at $q_\p \sim q^2_\perp/k$. Integral~\eqref{corrT0} can be estimated as
\be
  \delta S_0 \sim \int^k_0 d q_\perp \int^k_{q^2_\perp/k} d q_\p \frac{k}{q_\p} \sim k^2\ll1.
\ee
Then, zero-point oscillations in the considered system are rather weak.

In contrast, thermal correction~\eqref{corrT} has a power-law divergence that is stronger than the logarithmic one in isotropic 2D magnets. To see this, one notes that Planck's function at $T\gg\varepsilon_{\bf q}$ has the form  
\be
  \frac{1}{e^{\ve_\m{q}/T}-1} \approx \frac{T}{\ve_\m{q}}
\ee
so that from Eq.~\eqref{corrT} one has 
\be \label{corrT2}
  \delta S_T \sim \frac TA \int^k_0 d q_\perp \int^k_{q^2_\perp/k} \frac{d q_\p}{q_\p^2}.
\ee
In real systems, the power-law divergence in Eq.~\eqref{corrT2} can be screened by a small gap $\Delta$ in the spectrum which can arise, in particular, due to small low-symmetry spin interactions not considered in the present study. Moreover, the small gap should appear even in the considered model via the order-by-disorder mechanism after accounting for quantum fluctuations (see Ref.~\cite{Maleyev2006} for the corresponding discussion for cubic B20 helimagnets). Thus, in the classical limit, any finite temperature destroys the long-range magnetic order while quantum fluctuations can make the N\'eel temperature $T_N$ in the considered model finite by producing the gap $\Delta$ in the spectrum. Eq.~\eqref{corrT2} gives a simple estimation $T_N\sim\sqrt{\Delta A}$.

At $T>T_N$, due to the spectrum anisotropy, there are two correlation lengths $L_\|$ and $L_\perp$ of the short-range order in directions along and perpendicular to $\bf k$, respectively. According to Eq.~\eqref{spec5}, they are related as
\be \label{corrl0}
  L_\|\sim k L_\perp^2,
\ee
in order magnons with momenta corresponding to the correlation volume boundary have energies of the same order. Then, the correlation lengths can be found using the criterion $\delta S_T \sim S$ which reads as
\be \label{corrT3}
  \frac TA \int^k_{L^{-1}_\perp} d q_\perp \int^k_{\max\{L^{-1}_\|,q^2_\perp/k\}} \frac{d q_\p }{q_\p^2}
  \sim S.
\ee
Eqs.~\eqref{corrl0} and \eqref{corrT3} yield
\be\label{corl}
  L_\perp \sim \frac{SA}{T} \frac{1}{k},
  \qquad
  L_\| \sim \left(\frac{SA}{T}\right)^2 \frac{1}{k}.
\ee
which signify that $L_\perp\ll L_\|$ at $T\ll SA$ and $L_\|,L_\perp\gg1/k$.

To summarize the finite temperature estimations, one can see that (i) the correction to the staggered magnetization is a subject of power-law singularity, rather than a conventional logarithmic one; the latter usually allows to consider the so-called quasi-long-range order regime, which is not the case here (only short-range-order persists), and (ii) the coherent volume is essentially elongated in the modulation vector directions. The latter knowledge allows us to speculate on the spin ordering of the whole structure. Bearing in mind that in real materials small in-plane anisotropic interactions (e.g., anisotropic exchange) dictate preferable directions for cycloid modulation vectors $\m{k}$, we believe that at $T >T_N$ the system could be considered as a ``soup'' of elongated rectangular cycloid domains with two possible perpendicular orientations.

\section{Cycloid ordering and quenched disorder}
\label{SecDis}

\begin{figure}
  \centering
  \includegraphics[width=3.5cm]{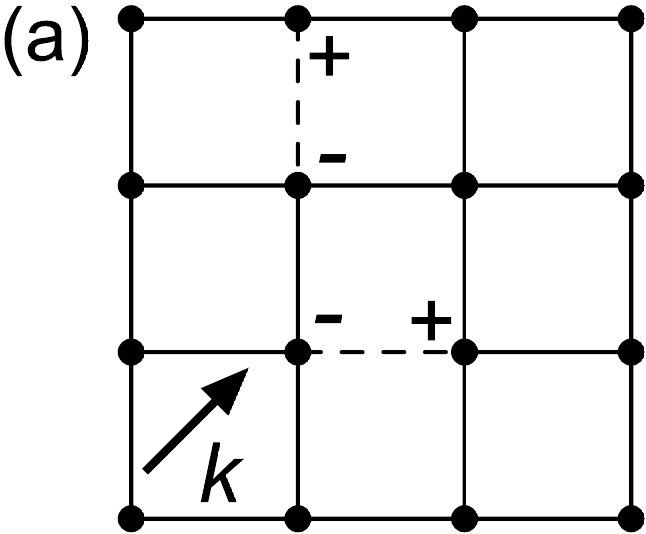} \hspace{1cm} \includegraphics[width=3.5cm]{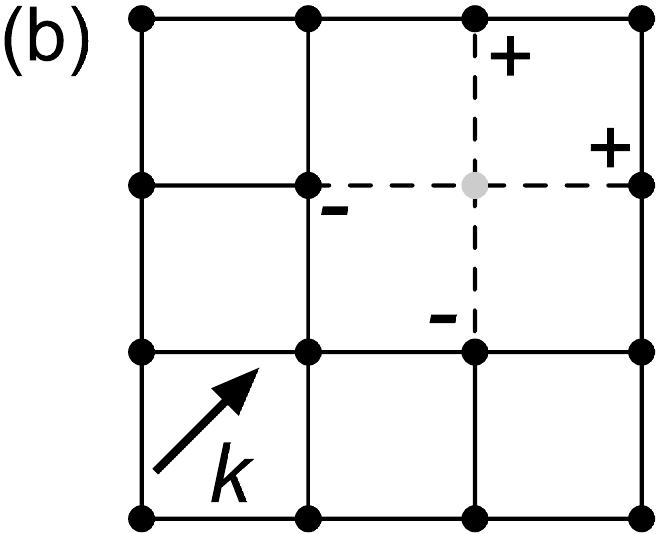}
  \caption{Two types of defects considered in the present study. (a) Defect bonds are shown by dashed lines which can be a result of some defect in the substrate. (b) Vacancy in the magnetic layer. In both cases, the distortion of the magnetic order by the defect is described by an equation for the field of electric charges indicated in both panels (see the text). }\label{FigDef}
\end{figure}

In this section, we consider the influence of quenched disorder on the magnetic ordering at $T=0$ in the classical limit. One expects various types of point-like defects on the interface in real systems. We consider two typical point defects in the magnetic monolayer shown in Fig.~\ref{FigDef}: (i) a defect bond with changed values of the interaction between neighboring magnetic ions which arises due to some impurity in the substrate, and (ii) a vacancy in the magnetic layer. We demonstrate that
even an infinitesimal concentration of such defects destroys the long-range order (previously, the same effect was discussed for frustrated noncollinear antiferromagnets in Ref.~\cite{santanu2020}). 

\subsection{Single defect} 

According to the linear response (Kubo) formalism~\cite{Kubo}, the defect-induced perturbation of the magnetic ordering $\delta {\bf S}_{\bf R}$ at site $\bf R$ reads as 
\be\label{dsgen}
\delta S^\alpha_{\bf R} =
\int \frac{d^2 q}{(2 \pi)^2} 
e^{i (\m{q} \cdot \m{R})}
\chi^{\alpha\beta}(\m{q}) {\cal Q}^\beta(\m{q}),
\ee
where $\chi^{\alpha\beta}(\m{q})$ is the tensor of static susceptibilities and $\mbox{\boldmath $\cal Q$} (\m{q})$ is the Fourier transform of the effective field produced by a defect.
It is clear from Eqs.~\eqref{SpinRep} that in the classical limit one needs transverse susceptibilities $\chi^{\eta \eta}(\omega,\m{q})$, $\chi^{\xi \xi}(\omega, \m{q})$, and $\chi^{\xi \eta}(\omega, \m{q})$ at $\omega\to0$. As one can see from Eq.~\eqref{SpinRep}, they are expressed via Green's functions $G_0$, $F_0$, and $F_0^\dagger$ introduced in Eqs.~\eqref{gf}. In the leading order in small parameters, $G_0$ is given by Eq.~\eqref{Gnormal} and one has 
\be \label{Ganomal}
  F_0(q) = F^\d_0(q) \approx - \frac{\tilde{\cal B}_\m{q} }{\omega^2 - \ve^2_\m{q}}
\ee
for the anomalous Green's functions, where $\tilde{\cal B}_\m{q}$ and $\ve_\m{q}$ are given by Eqs.~\eqref{b} and \eqref{spec5}, respectively. The non-diagonal static susceptibility $\chi^{\xi \eta}(\m{q})=0$. Diagonal static susceptibilities are approximately equal to each other at short wavelengths (as in a ferromagnet)
\be \label{SuscSh}
  \chi^{\eta \eta}(\m{q}) \approx \chi^{\xi \xi}(\m{q}) \approx \frac{S}{Aq^2}, \quad q \gg k,
\ee
whereas in the long-wavelength limit, they are essentially different:
\be \label{SuscL}
  \chi^{\eta \eta}(\m{q}) \approx S \frac{A k^2 + S B}{\ve^2_\m{q}}, \quad q \ll k, \\
  \chi^{\xi \xi}(\m{q}) \approx S \frac{A q^2}{\ve^2_\m{q}} \sim O(1), \quad q \ll k.
\ee
The latter equation indicates that the local transverse field in the $\xi$ direction (perpendicular to the plane in which spins rotate) would lead only to a local, visible on the scale of the cycloid period $\lambda = 2 \pi/k$, perturbation of the magnetic ordering. 

Let's consider a single defect in the substrate. It seems reasonable to model its influence on magnetic properties by a defect bond connecting neighboring magnetic atoms [see Fig.~\ref{FigDef}(a)]. We denote the deviation of the exchange coupling constant and iDMI at the defect bond from their values in the clean system as $U_J$ and $U_D$, respectively. We also introduce an explicit form for the cycloid modulation vector $\m{k} = k ( \cos{\t}, \sin{\t},0)$. The perturbation of the Hamiltonian \eqref{ham1} from the defect bond between spins at sites $\m{R}_0$ and $\m{R}_0 + \hat{x}$ reads as
\be \label{Def1}
  \mathcal{V} &=& U_D \hat{y} \cdot [\m{S}_{\m{R}_0} \times \m{S}_{\m{R}_0 + \hat{x}} ] - U_J \m{S}_{\m{R}_0} \cdot \m{S}_{\m{R}_0 + \hat{x}}\\
  &=& (k U_J - U_D) \cos{\t} \left( S^\zeta_{\m{R}_0} S^\eta_{\m{R}_0 + \hat{x}} - S^\eta_{\m{R}_0} S^\zeta_{\m{R}_0 + \hat{x}}  \right).\nn
\ee
Noteworthy, there are also terms including $S^\zeta S^\xi$ with coefficients $\propto \cos{\m{k} \m{R}_0}$, but they are oscillatory and fields along $\xi$ have only a local influence on the cycloid ordering. We also neglect terms containing an even number of Bose operators (e.g., $S^\zeta S^\zeta, S^\eta S^\eta$, etc.).

Bearing in mind that $S^\zeta_{\m{R}}\approx S$, the perturbation~\eqref{Def1} has the form of a Zeeman term produced by local magnetic fields directed along $\eta$ and acting on two spins at $\m{R}_0$ and $\m{R}_0 + \hat{x}$:
\be \label{Def2}
  \mathcal{V} &\cong&  {\cal Q} \cos{\t} \left( S^\eta_{\m{R}_0} - S^\eta_{\m{R}_0 + \hat{x}} \right),\\
  \label{f}
  {\cal Q} &=& S (U_D - k U_J).
\ee
Substituting these expressions to Eq.~\eqref{dsgen}, one obtains
\be
\label{DefField}
  \delta S^\eta_\mathbf{R} &=& {\cal Q} \cos{\t} \\ &&\times \int \frac{d^2 q}{(2 \pi)^2} \chi^{\eta \eta}(\m{q}) 
  \left[e^{i \m{q} \cdot (\m{R} - \m{R}_0)} - e^{i \m{q} \cdot (\m{R} - \m{R}_0 - \hat{x})} \right]. \nn
\ee
At short distances $|\m{R} - \m{R}_0| \lesssim \lambda$, the integral in Eq.~\eqref{DefField} is mainly determined by $q \gtrsim k$ so that $\chi^{\eta \eta}$ should be taken from Eq.~\eqref{SuscSh}. Then, Eq.~\eqref{DefField} gives the field of two Coulomb charges (a dipole) in the two-dimensional space having the form 
\be
  \delta S^\eta_\mathbf{R} = \frac{S{\cal Q} \cos{\t}}{2 \pi A} \Bigl(
  \ln{\left|\m{R} - \m{R}_0 - \hat{x}\right|} - \ln{\left|\m{R} - \m{R}_0\right|}
  \Bigr).
\ee
This expression is simplified as follows at $1\ll\left|\m{R} - \m{R}_0 \right|\alt\lambda$:
\be\label{dssim}
  \delta S^\eta_\mathbf{R} = - \frac{S{\cal Q} \cos{\t}}{2 \pi A} \frac{(\m{R} - \m{R}_0) \cdot \hat{x}}{(\m{R} - \m{R}_0)^2}.
\ee
Then, the defect bond behaves as an electric dipole. Such electrostatic analogy was previously exploited in Refs.~\cite{villain1977two,villain1979insulating,aharony,korenblit,utesov2015,utesov2019,syromyatnikov2021} in disordered magnets with isotropic long-wavelength spectrum. It is easy to show that the defect bond between sites $\m{R}_0$ and $\m{R}_0 + \hat{y}$ leads to the similar result
\be
  \delta S^\eta_\mathbf{R} = \frac{S{\cal Q} \sin{\theta}}{2 \pi A} \Bigl(
  \ln{\left|\m{R} - \m{R}_0 - \hat{y}\right|} - \ln{\left|\m{R} - \m{R}_0\right|}
  \Bigr).
\ee

At large distances $|\m{R} - \m{R}_0| \gg \lambda$, momenta $q \ll k$ dominate in Eq.~\eqref{DefField} due to the oscillatory factor. In this case, the susceptibility $\chi^{\eta\eta}(\m{q})$ is essentially anisotropic as a function of momentum. As a result, the cycloid distortion $\delta S^\eta_\m{R}$ at large distances is no longer the field of the 2D dipole.
We set $B=0$ below for simplicity since a finite $B$ does not affect the physical picture. We obtain using Eqs.~\eqref{SuscL} and~\eqref{DefField}
\be
   \delta S^\eta_\mathbf{R} &=& S {\cal Q} \cos{\t} \int_{-\infty}^\infty \frac{dq_\|dq_\perp}{(2 \pi)^2} \frac{k^2}{A(k^2 q^2_\| + g^2q^4_\perp)}\nn\\
   &&\times\left[ e^{i \m{q} \cdot (\m{R} - \m{R}_0)} - e^{i \m{q} \cdot (\m{R} - \m{R}_0 - \hat{x})} \right]. 
\ee
After integration by residues over $q_\|$ one has
\be \label{ds1}
   \delta S^\eta_\mathbf{R} &=& \frac{S k {\cal Q} \cos{\t} }{4\pi g A} \\
   &\times& \int_{-\infty}^\infty \frac{dq_\perp}{q_\perp^2}\left( e^{i q_\perp R_1^\perp - g q_\perp^2 R_1^\|/k} - e^{i q_\perp R_2^\perp - g q_\perp^2 R_2^\|/k} \right), \nn
\ee   
where
\be
   R_1^\perp &=& (\m{R} - \m{R}_0)\cdot\hat c, \\
   R_1^\| &=& |(\m{R} - \m{R}_0)\cdot{\bf k}|/k,\\
   R_2^\perp &=& (\m{R} - \m{R}_0 - \hat x)\cdot\hat c,
   \\
   R_2^\| &=& |(\m{R} - \m{R}_0-\hat x)\cdot{\bf k}|/k.
\ee
The integral in Eq.~\eqref{ds1} can be taken analytically with the result
\be\label{ds2}
   \delta S^\eta_\mathbf{R} &=& \frac{S k {\cal Q} \cos{\t} }{4 g A} \left\{
   \frac{2}{\sqrt\pi}\Biggl[
   \sqrt{\frac{g R_2^\|}{k}} \exp{\left( -\frac{k R_2^{\perp2}}{4 g R_2^\|} \right)} \right. \nn \\
   && {}-  \sqrt{\frac{g R_1^\|}{k}} \exp{\left( -\frac{k R_1^{\perp2}}{4 g R_1^\|} \right)}
   \Biggr] \\
   &&{}+ \left.
   R_2^\perp{\rm Erf}\left(\frac{\sqrt{k} R_2^\perp}{2\sqrt{g R_2^\|}}\right)
   -   
   R_1^\perp{\rm Erf}\left(\frac{\sqrt{k} R_1^\perp}{2\sqrt{g R_1^\|}}\right)
   \right\},\nn
\ee
where ${\rm Erf}(x)$ is the error function. Importantly, in a relatively wide region of $k (R^\perp_i)^2 \agt |R^\p_i|$, where $i=1,2$,  Eq.~\eqref{ds2} has a very simple form in the considered regime of $|\m{R} - \m{R}_0| \gg \lambda$:
\be \label{ds3}
  \delta S^\eta_\mathbf{R} \cong -  \frac{S k {\cal Q} \cos{\t} \sin{\t} }{4 g A} \textrm{sign}\left[(\m{R} - \m{R}_0)\cdot\hat c \right]
\ee
with corrections being of the order of $\exp{(- k |\m{R} - \m{R}_0|})$. In another limiting case of $k (R^\perp_i)^2 \ll |R^\p_i|$ one has
\be \label{ds4}
  \delta S^\eta_\mathbf{R} &=& -  \frac{S {\cal Q} \cos{\t}^2 }{4 A} \sqrt{\frac{k}{g |\m{R} - \m{R}_0|}} \textrm{sign}\left[(\m{R} - \m{R}_0)\cdot \m{k} \right] \nn \\ && {} + O\left(1/|\m{R} - \m{R}_0|^{5/2}\right).
\ee
It is easy to show that in the case of the defect bond directed along $\hat{y}$ axis one should simply make a substitution $\t \mapsto \t - \pi/2$ in Eqs.~\eqref{ds3} and~\eqref{ds4}.

\begin{figure}
  \centering
  \includegraphics[width=8cm]{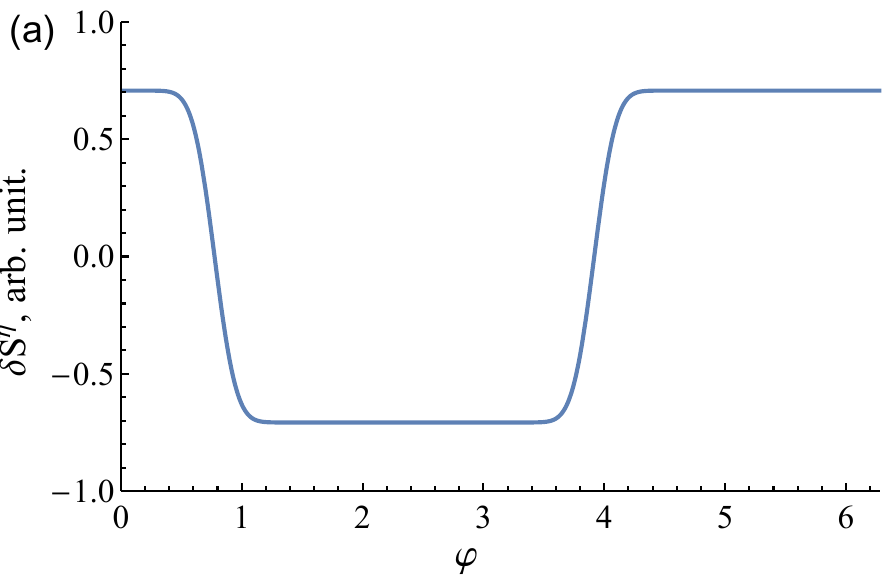} \hspace{1cm} \includegraphics[width=8cm]{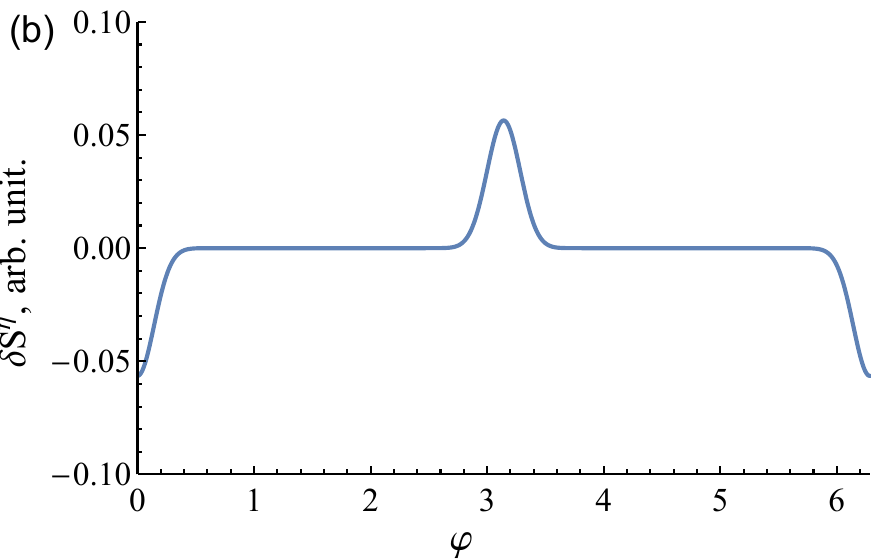}
  \caption{The cycloid distortion $\delta S^\eta_\mathbf{R}$ given by Eq.~\eqref{ds2} and produced by a single defect bond between $\m{R}_0$ and $\m{R}_0 +\hat{x}$ sites at  large distances $|\m{R} - \m{R}_0| \gg \lambda$. The dependence is shown of $\delta S^\eta_\mathbf{R}$ on the polar angle $\varphi$ of $\m{R} - \m{R}_0$ for a certain set of the model parameters. The cycloid modulation vector $\bf k$ is oriented (a) along the square diagonal and (b) parallel to $\hat{x}$ axis. }\label{FigField}
\end{figure}

Fig.~\ref{FigField} illustrates the behavior of $\delta S^\eta_\mathbf{R}$ given by Eq.~\eqref{ds2}. When $\m{k}$ is not oriented along square edges, asymptotic \eqref{ds3} works at all orientations of ${\bf R}-{\bf R}_0$ except for close vicinity of directions parallel to ${\bf k}$. In this case, $\delta S^\eta_\mathbf{R}$ is described by the field of a 1D dipole whose axis is oriented parallel to $\hat{c}$ [see Fig.~\ref{FigField}(a)]. In contrast, when $\m{k}$ is directed along a square edge, the prefactor in Eq.~\eqref{ds3} is zero and $\delta S^\eta_\mathbf{R}$ acquires a small value only when the condition of applicability of asymptotic~\eqref{ds4} holds, i.e., when ${\bf R}-{\bf R}_0$ is nearly parallel to ${\bf k}$. Note that the distortion made by the defect along the $\hat{y}$ axis is exactly zero in this case.

The vacancy shown in Fig.~\ref{FigDef}(b) does not require a separate consideration because it can be treated as a superposition of four defect bonds with, e.g., $U_D = - D$ and $U_J = -J$ in the nearest-neighbors model. At short length scales, it leads to a dipole-like $\delta S^\eta_\mathbf{R}$ with the dipole momentum directed along $\m{k}$. At large distances, it is easy to show that constant contributions to $\delta S^\eta_\mathbf{R}$ given by Eq.~\eqref{ds3} cancel each other at any orientation of $\m{k}$. In the main order, the result in this case is similar to that shown in Fig.~\ref{FigField}(b), where a small cycloid distortion arises only when ${\bf R}-{\bf R}_0$ is nearly parallel to ${\bf k}$.


\subsection{Finite concentration of defects}

We turn to the discussion of a finite concentration $c$ of defects. Let us start with the case of defect bonds. We assume that the probability for each particular bond to be the defect one is equal to $c \ll 1$. Various types of disorder can be studied. We concentrate below on the binary one at which all the defect bonds are characterized by the same parameters. The theory adaptation for other types of disorder, e.g., for the Gaussian one, is straightforward.

It was shown in Ref.~\cite{santanu2020} that even an infinitesimal concentration of defect bonds is sufficient to destroy the long-range order in frustrated non-collinear 2D helimagnets. In simple models like triangular-lattice antiferromagnet discussed in Ref.~\cite{santanu2020}, spin textures produced by defect bonds are described by expressions like Eq.~\eqref{dssim}. Then, the spin at a certain lattice site feels dipole-like fields from all the defect bonds. After averaging over disorder configurations, the mean square transverse spin component in the linear order in defects concentration $c$ can be estimated as 
\be\label{dsp}
  \langle \delta S^2_\perp \rangle \sim c \int d^2 \m{r} \left( \frac{\m{d} \cdot \m{r} }{r^2}\right)^2 \sim c d^2 \int_1^\infty \frac{d r}{r},
\ee
where $\m{d}$ is the defect bond dipole moment. 
The integral in Eq.~\eqref{dsp} diverges at large distances. Then, one needs to introduce a cut-off whose physical meaning is the disorder-induced correlation length. It can be estimated using the condition $\langle \delta S^2_\perp \rangle \sim S^2$,  which yields
\be \label{corr2D}
  L_d \sim e^{{\cal G}/c d^2},
\ee
where ${\cal G}$ is a constant. For $c \ll 1$, the correlation length is macroscopically large and a quasi-long-range order persists.

The situation is essentially different in the magnet with the iDMI. At $c \ll 1$, the most important part of the cycloid distortions comes from large lengths, where, in general, we should use quasi-1D solution \eqref{ds3}. The condition 
$\langle \delta S^2_\perp \rangle \sim S^2$ reads as $c(Sk{\cal Q}/A)^2L_d^2\sim S^2$ and gives
\be \label{Ld1}
  L_d \sim \frac{A}{{\sqrt{c} |\cal Q}|} \frac{1}{k}.
\ee
Then, instead of the exponential dependence on $c$, we find the power-law one so that the effect of disorder is much more pronounced in the considered system.

In the case of vacancies or defect bonds, when $\m{k}$ is parallel to a square edge, 
we have to use Eq.~\eqref{ds4} for estimations. Bearing in mind that the corresponding contribution is nonzero only at a small interval $\sim 1/\sqrt{k |{\bf R}-{\bf R}_0|}$ of directions of ${\bf R}-{\bf R}_0$ along $\bf k$, we write
\be
  \langle \delta S^2_\perp \rangle \sim c \frac{S^2 {\cal Q}^2}{A^2} \int r dr \frac{\sqrt{k}}{r^{3/2}}.
\ee
Then, the disorder-induced correlation length is estimated as 
\be \label{Ld2}
  L_d \sim \frac{A^4}{ {c^2 \cal Q}^4} \frac{1}{k}
\ee
whose dependence on the defects concentration is still the power-law one rather than the exponential one. Notice that in contrast to the considered system, an infinitesimal concentration of vacancies and other defects with a symmetric arrangement of defect bonds do not destroy the long-range order in simple spiral magnets like triangular-lattice antiferromagnet. \cite{santanu2020,syromyatnikov2021}

The way in which results~\eqref{Ld1} and~\eqref{Ld2} are obtained suggests that the same equations should hold for other types of disorder rather than the considered binary one. For example, in the case of some continuous distribution of ${\cal Q}$ among defects, we should simply substitute ${\cal Q}^2$ with its average value $\langle {\cal Q}^2 \rangle$ in Eqs.~\eqref{Ld1} and~\eqref{Ld2}. No qualitative changes in the results are expected. In any case, an infinitesimal amount of quenched disorder should lead to short-range cycloid structures even at zero temperature.

\section{Discussion and conclusion}
\label{SecConc}

To conclude, we consider the ferromagnetic monolayer with interfacial Dzyaloshinskii-Moriya interaction (iDMI) and a small easy-axis anisotropy \eqref{ham1}. This model should also be applicable to thin films with a strong enough iDMI. The cycloid magnetic order with several possible ordering vectors $\bf k$ arises in this system that makes possible also a topological magnetic structure. In the present study, we analyze only the state with the cycloid magnetic ordering characterized by a single ordering vector $\bf k$ with $k\ll1$. 

We derive the classical spectrum of elementary excitations (magnons) at zero temperature. We show that the spectrum of the short-wavelength magnons is similar to that in a common ferromagnet. In contrast, umklapp terms in the bosonic version of the spin Hamiltonian renormalize the long-wavelength spectrum drastically making it essentially anisotropic with respect to the $\m{k}$ direction: it is linear and quadratic for momenta directed along and perpendicular to $\bf k$, respectively. Then, the spin-wave velocity of this quasi-1D spectrum is essentially angle-dependent so that it is zero in the direction transverse to $\bf k$. 

Umklapps also lead to the emergence of magnon bands inside the Brillouin zone similar to electronic bands in solids. We show that at $q_\perp \neq 0$ noticeable gaps in the spectrum emerge at $q_\| = \pm k/2, \pm 3k/2,\dots$ due to iDMI single umklapps and at $q_\| = \pm k,\pm2k,\dots$  due to the anisotropy-induced double umklapps, where $q_\perp$ and $q_\|$ are components of the momentum $\bf q$ perpendicular and parallel to $\bf k$, correspondingly (see Figs.~\ref{FigSpec} and \ref{FigSpec2}). Moreover, the magnon dispersion inside the bands becomes almost flat in $q_\p$ at $q_\perp \sim k$ as it is seen in Fig.~\ref{FigSpec2}. This magnon band structure appears in dynamical spin correlators as a sequence of numerous anomalies which can be observed experimentally (see Fig.~\ref{FigStruct2}).

The obtained quasi-1D character of the spectrum has important consequences on the physical properties of the considered system on large-length scales. At $T=0$, the quantum correction to the mean magnetic moment $\langle m\rangle$ is small ($\sim k^2\ll1$) and the cycloid order is robust. However long-wavelength thermal fluctuations produce a correction to $\langle m\rangle$ with a power-law singularity rather than the logarithmic one (which is usual for isotropic 2D magnets). Thus, even so-called quasi-long-range order cannot persist in the considered system at finite $T$. Our estimations show that correlation lengths are substantially different at $T\ne0$ for directions along $\bf k$ and perpendicular to it. Thus, the coherent volume of the short-range order is substantially elongated in the $\m{k}$ direction.

We also consider two typical types of impurities in the magnetic layer which can arise in real systems: defect bonds and vacancies. In both cases, we arrive at qualitatively similar conclusions that a single defect provides long-range perturbations of the cycloid structure which destroys the long-range order even at $T=0$ at finite defects concentration $c\ll1$. Importantly, we obtain that the cycloid distortion caused by a single impurity is described by an electrostatic field produced by a complex of charges (one and four 1D dipoles for the defect bond and the vacancy, respectively) placed at the defect location (see Fig.~\ref{FigDef}). As a result, the correlation length in the impurity-induced disordered ground state has strong power-law dependencies on $1/c$.

We would like to note that in real systems there are several possibilities for screening the discussed above large-length-scale peculiarities. 
First, small spin interactions arising in the fourth order in the spin-orbit coupling can lead to a small velocity of spin waves at $q_\p =0$. Second, quantum corrections should lead to the gap in the magnon spectrum via the order-by-disorder mechanism. These corrections introduce an extra very long length scale (or, equivalently, low-energy) parameter to the theory up to which the above consideration remains valid. 

\begin{acknowledgments}

This work is supported by the Russian Science Foundation (Grant No. 22-22-00028). O. I. U. acknowledges financial support at the final stage of this research from the Institute for Basic Science in Korea (Project No. IBSR024-D1).

\end{acknowledgments}

\bibliography{TAFbib}

\end{document}